# DNA origami assembled nanoantennas for manipulating single-molecule spectral emission

*María Sanz-Paz[1§*], Fangjia Zhu[1§], Nicolas Bruder[1], Karol Kołątaj[1], Antonio Fernández-Domínguez[2*], and Guillermo P. Acuna[1*]*

[1]Department of Physics, University of Fribourg, Chemin du Musée 3, Fribourg CH-1700, Switzerland.

[2] Departamento de Física Teórica de la Materia Condensada and Condensed Matter Physics Center (IFIMAC), Universidad Autónoma de Madrid, E-28049 Madrid, Spain.

§These authors contributed equally.

**Abstract:** Optical nanoantennas can affect the decay rates of nearby emitters by modifying the local density of photonic states around them. In the weak-coupling limit, and according to the Fermi's Golden Rule, the emission spectrum of a dye is given by the energy of all the possible radiative transitions weighted by the probability of each of them to occur. By engineering the resonance of a nanoantenna, one can selectively enhance specific vibronic transitions of a dye molecule, thus shaping its emission spectrum. Since interactions between emitters and nanoantennas are known to be position dependent, we make here use of DNA origami to precisely place an individual dye at different positions around a gold nanorod. We show how this relative position between the nanorod and the emitter affects the emission spectrum of the latter. In particular, we observe the appearance of a second fluorescence peak whose wavelength and intensity are correlated with the fundamental plasmonic resonance of the nanorod, which we extract from its photoluminescence spectrum. This second peak results from the selective enhancement of transitions to different vibrational levels of the excitonic ground state, whose energies are in resonance with the plasmonic one. Furthermore, we argue that the drastic alteration of the fluorescence spectrum in some of our samples cannot be accounted for with Kasha's rule, which indicates that radiative and vibrational relaxation dye lifetimes can become comparable through the coupling to the gold nanorods.

Plasmonic nanostructures can act as optical antennas[1] capable of mediating interactions between light and fluorescent emitters.[2] For example, they can couple free-space propagating light into highly enhanced and localized near-fields, and therefore increase the excitation rate of adjacent emitters.[3] In addition, even in the weak-coupling regime, plasmonic nanostructures can also tailor the emission process, as they can modulate the local density of states (LDOS)[4,5] around them, thus affecting the decay rates of nearby emitters according to Fermi's Golden Rule.[6] These changes in the radiative and nonradiative decay rates influence in turn the photophysical properties of such fluorescent emitters,[7] that can experience strong fluorescence (FL) enhancement,[3,8] lifetime shortening,[9] enhancement/reduction of their quantum yield,[10] increment of their photostability,[11,12] changes in their emission pattern,[13–15] and a shift of their apparent emission center.[16] Furthermore, LDOS experienced by single emitters are known to be dependent on their distance and orientation to the plasmonic structure[17] as well as on their spectral overlap.[18,19]

In most cases, the phenomena above could be interpreted in terms of the Purcell effect, in which dye molecules were modelled as two-level systems. In this way, they are described by a single, sharp emission wavelength, which arises solely from the transition between two electronic/excitonic levels. Although this model can successfully explain multiple experimental observations, the energy states of dye molecules are actually more complex, since both ground and excited electronic states have multiple vibrational levels associated.[20] As illustrated in Figure 1A, the emission spectrum of a dye molecule is given by the energy of the possible vibronic transitions to its ground state, weighted by their respective transition probabilities (the so-called Franck-Condon factors).[21] In the presence of an optical nanoantenna, this spectrum can be reshaped via selectively enhancing some specific radiative transitions through the plasmon resonance. This is due to the wavelength-dependent radiative rate enhancement (given by the LDOS) produced by the nanoantenna, and can only be explained considering the multilevel nature of the dye molecule.[22] This plasmon-induced spectral reshaping can be exploited in a variety of applications including the plasmonic engineering of the emission colors of light-emitting devices,[23] or for tailoring the spectral overlap in dipole-dipole energy transfer processes.[24]

To date, a handful of works have investigated this effect, albeit either on experiments at the ensemble level[22,25,26] or in the presence of diffusing molecules at reduced concentrations.[27] These pioneering examples constitute the first evidence of the potential of nanoantennas to manipulate the emission spectra of nearby dye molecules. In order to acquire a deeper understanding of this effect, that will unlock a higher level of emission manipulation, it is necessary to perform experiments at the single molecule level with a controlled distance and relative position between the emitter and the nanoantenna. Here, using the DNA origami technique,[28–31] we investigate plasmon-assisted emission reshaping with an unprecedented control, in the nm range, of the distance between an individual dye and a nanoantenna based on a single colloidal gold nanorod (AuNR). We show that the FL spectrum of a single dye molecule coupled to a AuNR can be greatly modified, and that the degree of reshaping depends on the strength of both the spectral and spatial overlap with the plasmon resonant fields. Since such spectral variation occurs due to the selective enhancement of radiative transitions, we also demonstrate that we can experimentally retrieve the wavelength-dependent behavior of the radiative decay rate enhancement taking place in our hybrid dye-AuNR system. Moreover, we find signatures that indicate that photon emission lifetimes approach vibrational relaxation timescales in some of our samples, a regime in which fluorescence does not obey Kasha's rule.[20]

To fabricate the dye-AuNR samples, we used a T-shaped DNA origami structure[14] (Figure 1B) as a template to "host" previously functionalized colloidal AuNRs through DNA hybridization (further details are included in Figure S4). The origami design enables the incorporation of a single fluorescent molecule (ATTO 594) at the tip of the AuNR, 5 nm away from the gold surface (red dot in Figure 1B, placed at Pos1). We used commercial colloidal AuNRs with a nominal diameter and length of 40 nm and 80 nm, respectively. After assembly, the hybrid dye-AuNR structures were dried on a glass substrate and measured with a custom-built confocal scanning fluorescence microscope, where light from the center of the objective can be directed to a spectrometer (see experimental details in Methods).

Initially, we determined the emission spectrum of the single ATTO 594 dye within the DNA origami but in the absence of the AuNR for referencing (Figure 1C shows an average spectrum of more than 50 dye molecules). It exhibits a main peak at ~615 nm and a "shoulder" from a secondary transition at ~655 nm. Then, we proceeded by measuring the spectra of the dye-AuNR structures. This was carried out by recording time-dependent spectra, with an integration time of 1 second. From these measurements, FL intensity transients were generated. Only transients with a single bleaching step, a signature of emission arising from a single molecule, were employed for further analysis (see exemplary FL transient in Figure 1D). This approach, besides confirming that only one molecule was driving the antenna, enabled the discrimination of the single-molecule FL emission from the photoluminescence (PL) of the AuNRs,[14] by subtracting the latter to the total signal measured before bleaching. Furthermore, the PL spectrum is governed, for AuNRs with a rather low aspect ratio such as the ones employed in this work, by a single peak close to the longitudinal plasmonic resonance[32–34] (whose wavelength redshifts with increasing nanorod aspect ratio[35]). Since commercial AuNRs exhibit considerable size inhomogeneity (see Figure S1), the PL analysis enables a comprehensive characterization of the plasmonic response directly affecting the emission properties of the nearby dye, under the same sample and measurement conditions (see PL peak histogram plot in Figure 1C). The AuNRs size distribution, with the corresponding plasmonic resonance distribution, has therefore been exploited to investigate different spectral overlaps between the dye emission and the AuNR plasmonic response (Figure 1C). Finally, this analysis was combined with scanning electron microscopy (SEM) imaging to ensure that all structures studied contained only single AuNRs.

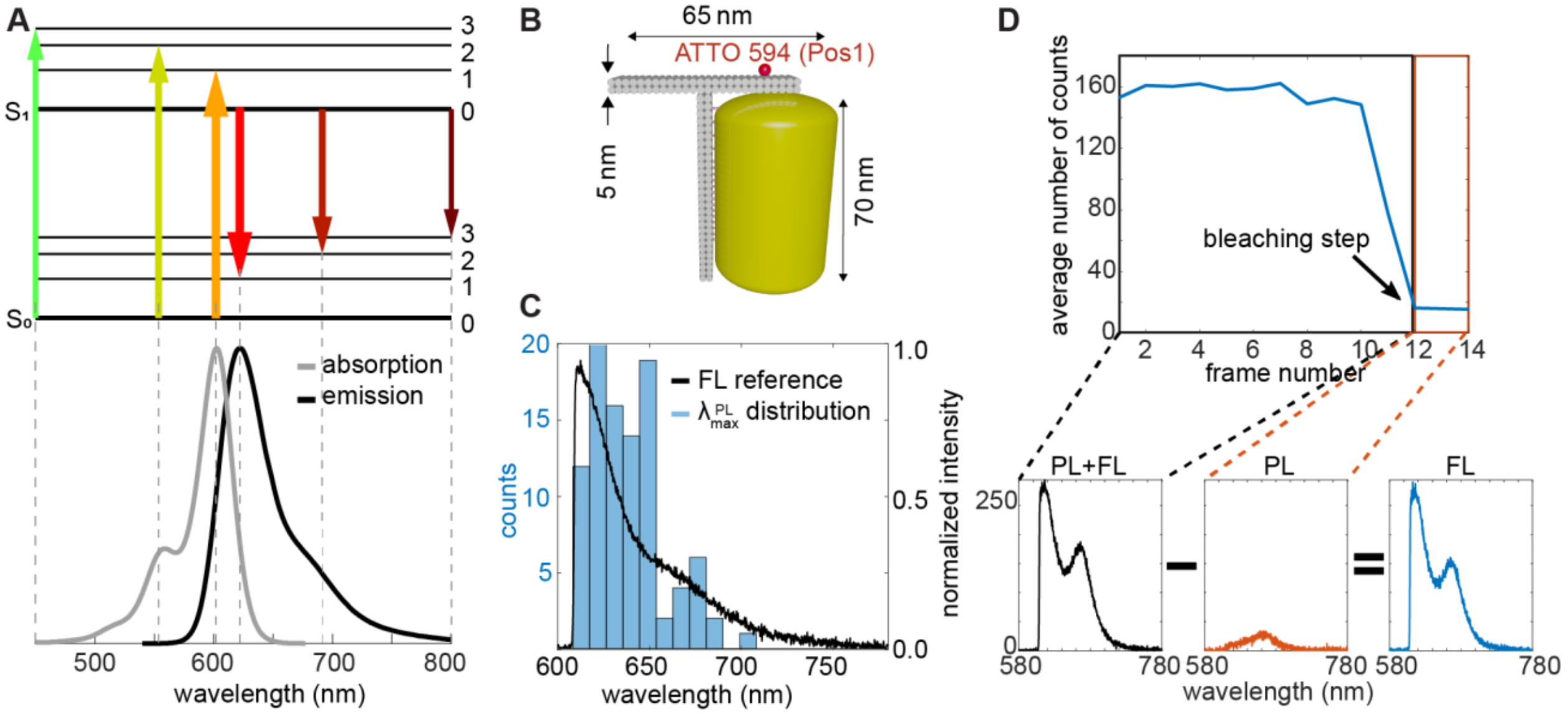

***Figure 1. Fluorescence spectrum of an ATTO 594 dye, DNA origami antenna geometry and analysis pipeline***. *(A) Simplified Jablonski diagram showing some excitation and emission transitions defining the absorption (grey) and emission (black) spectra of the ATTO 594 dye (data taken from atto-tec.com for a dye in PBS solution). (B) Sketch of the DNA origami structure hosting a AuNR and a single dye molecule (red ball) placed at Pos1. (C) Measured emission spectra of ATTO 594 (black line) in a surface-immobilized DNA origami in air, and histogram plot of the PL peak position of the AuNRs for all the structures analyzed. The spectrum is blue-shifted with respect to the one in (A) due to a change in the environment (experiments are done in air whereas reported spectrum from ATTO-TEC were measured in solution). (D) Exemplary fluorescence*

*transient employed to isolate the fluorescence (FL) and photoluminescence (PL) spectra by subtraction of the latter after bleaching.*

Figure 2 depicts both the FL and PL from the DNA origami-based dye-AuNR single structures, together with the normalized FL reference (dye in the absence of AuNRs) for comparison. For better visualization, spectra are separated in two columns according to the AuNR's PL. Two different scenarios can be clearly identified in terms of FL emission reshaping. In the first one, the AuNR PL overlaps with the main vibronic transition or is located between the main and the secondary (shoulder) FL peaks (PL peak < 630 nm, Figure 2A). In this case, two main effects are observed: the suppression of the shoulder that corresponds to the secondary vibronic transition and a displacement of the main FL peak to a position that coincides with the AuNR's PL. In the second one, the AuNRs present the PL maximum in the vicinity of the shoulder of the FL emission (PL peak > 630 nm, Figure 2B). In this situation, we observe the appearance of a second FL peak in the spectral range of the bare shoulder and AuNR peak, whose intensity can be higher or lower than the main FL peak. Furthermore, in some cases we even observe the suppression of the main FL peak.

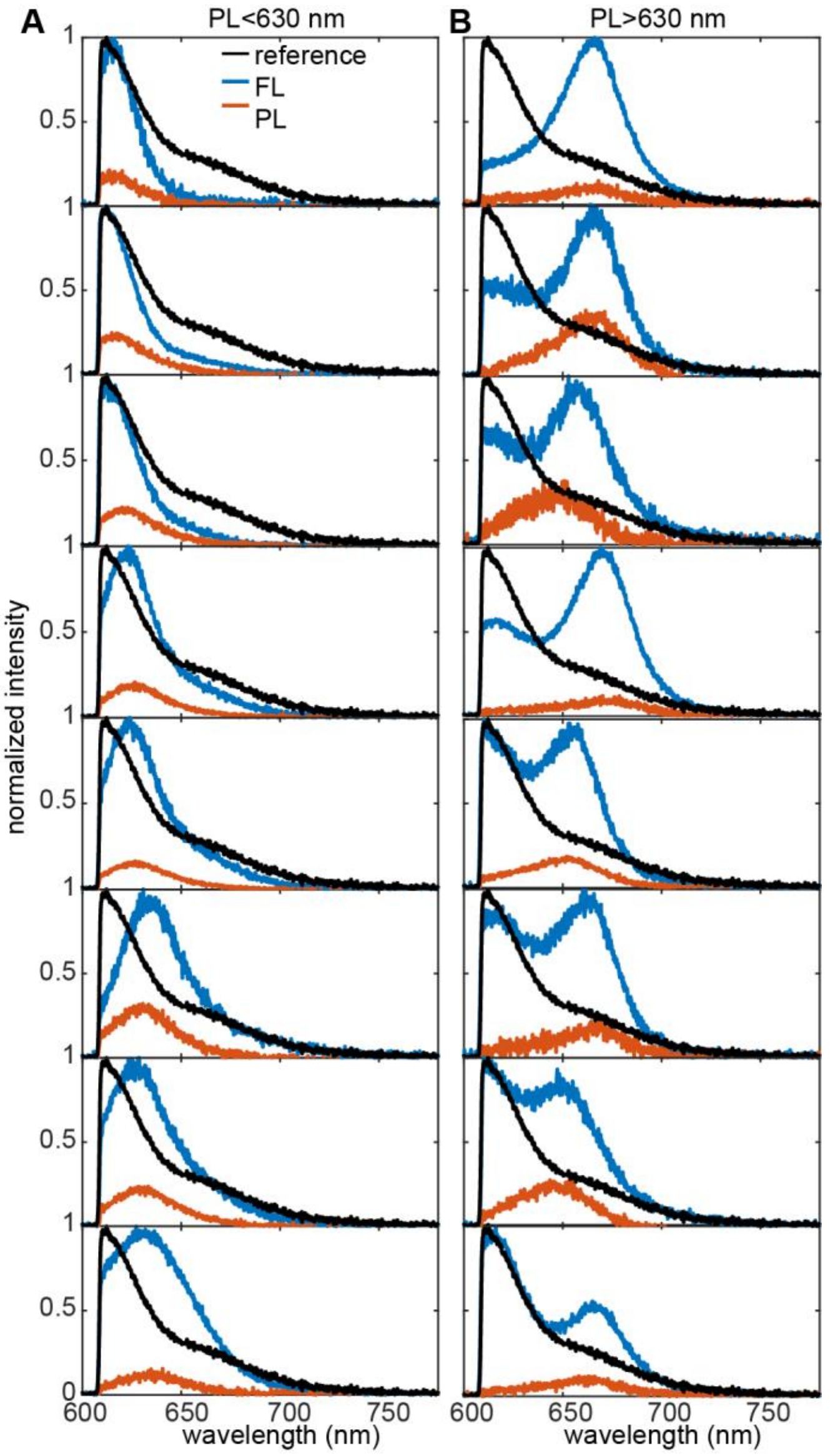

***Figure 2. Reshaped fluorescence spectra from dye-AuNRs with different PL.*** *(A,B) Fluorescence emission spectra for the dye without AuNR (reference, black line) and for the hybrid dye-AuNR structures (dye at Pos1). The AuNR PL is also included for each case (orange line). Spectra are divided into cases, with PL < 630 nm (A) and PL > 630 nm (B), i.e. in resonance with main and secondary (shoulder) FL peaks. PL spectra are normalized to the corresponding FL maxima for visualization of their intensity difference.*

In order to quantify the observed spectral reshaping, for each FL spectrum we determined both the position and the amplitude of the peaks. To do so, we fit the FL spectra to a sum of two Lorentzian functions and extract their peak amplitude and position. For the cases in which either the main or the secondary (shoulder) peaks are heavily suppressed, the FL spectra is fitted to a single Lorentzian function. For all cases, the PL spectra is also fitted with a single Lorentzian. First, we compare the position of the Lorentzian(s) fitted from the reshaped FL spectra versus the PL peak in Figure 3A. When only one Lorentzian is found (top panels of Figure 2A), we observe a predominately linear correlation between FL and PL peaks (linear fit with slope ~0.99). For the cases where two FL peaks appear (Figure 2B), the one with the shortest wavelength is always at the intrinsic position regardless of the PL, whereas the other one follows the spectral position of the AuNR PL peak (linear fit with slope ~1.01), in agreement with previously published results at the ensemble level.[22,25,26]

To compare the relative amplitude of the observed peaks, we defined the transition ratio $R$ as:[27]

$$R = \frac{I_1}{I_2} \frac{1}{R_{ref}} \tag{1}$$

where $I_1$ and $I_2$ are the amplitudes of the two fitted Lorentzian functions, and $R_{ref}$ is the amplitude ratio of the main and "shoulder" peaks in air and in the absence of AuNR, whose spectrum is obtained by averaging multiple dyes (Figure 1C). For the cases when the fitting of the reshaped FL yields only one Lorentzian, its amplitude defines $I_1$ ($I_2$) if its position is below (above) ~630 nm, and for $I_2$ ($I_1$) we use the amplitude at the wavelength where the secondary (main) transition in the reference spectrum is. This definition of the transition ratio implies that $R$ would be equal to 1 if the relative amplitudes remain unchanged with respect to the isolated molecule, while $R >$ 1 indicates an enhancement of the main transition over the secondary one, and $R < 1$ an enhancement of the secondary transition.[27]

The transition ratio $R$ was calculated for all of the structures measured, and the values obtained are displayed in Figure 3B. For the AuNRs in resonance with the main FL peak (PL peak <630 nm), most values lie above 1 and reach a maximum of 6.2, meaning that there is an enhancement of the main transition over the secondary one compared to the reference, which can also be seen by the suppression of the "shoulder" observed in the reference spectrum (Figure 2A). On the contrary, for AuNRs with a PL peak > 630 nm, we observe a large amount of structures showing $R < 1$. This implies an enhancement of secondary vibronic transitions, in agreement with published results obtained from diffusing molecules.[27] In particular, values as low as $R = 0.016$ were obtained, which means that the second peak (transition) overpasses the main one ($R = 0.307$ corresponds to both transitions being of equal strength) in intensity. Overall, this represents a 360-fold tuning of the relative intensities of the observed vibronic transitions with respect to the spectrum of an isolated molecule (6-fold for $R > 1$ times 60-fold for $R < 1$). We observe the enhancement of this second peak ($R < 1$) in more than 80% of the samples where the PL peak > 630 nm. Furthermore, the correlation between the PL peak and the $R$ values is apparent in Figure 3B. The more redshifted the PL is with respect to the main FL transition, the stronger the effect on the enhancement of the secondary transition (lower $R$). Figure 3B also shows a considerable dispersion in the correlation. This can be due to multiple factors, including the orientation of the dye and AuNR size inhomogeneity (Figure S1).[25] Altogether, this means that the resonant wavelength of the AuNRs defines not only the existence and position of the second FL peak, but also its relative intensity. This proves that AuNRs can selectively enhance specific radiative transitions that are in resonance with the plasmonic peak, thus reshaping the overall emission spectra of nearby, single dye molecules.

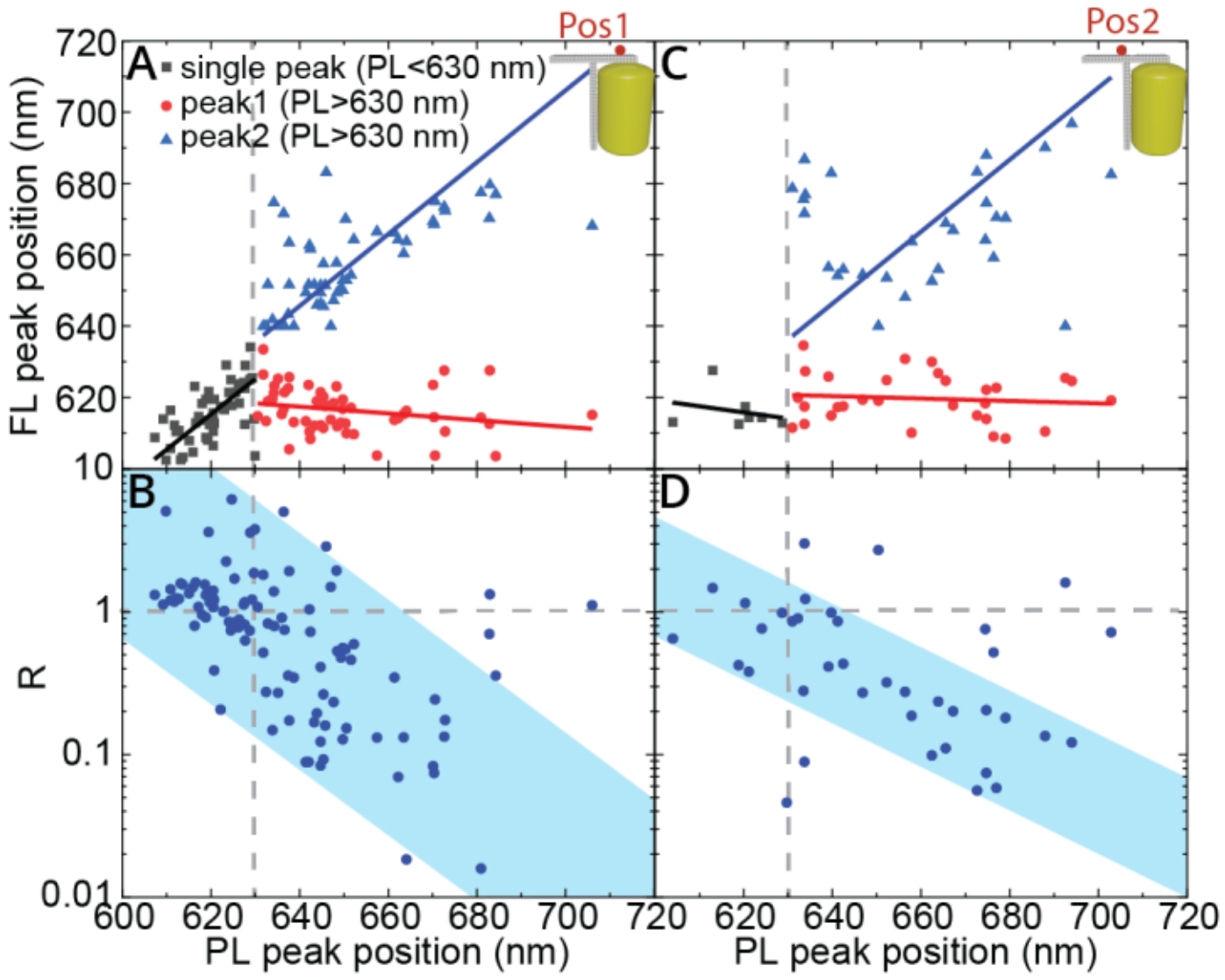

***Figure 3. Degree of fluorescence reshaping as a function of the AuNR PL.*** *(A,C) FL vs PL peak positions for two different locations of the dye with respect to the AuNR (Pos1 (A) and Pos2 (B) as shown in the insets). Data is divided into spectra with mostly one FL peak (black squares) and mostly two peaks (red dots and blue triangles), whose boundary is indicated by a vertical gray dashed line. Lines represent linear fits for each case, showing that for PL > 630 nm the position of the main peak is not affected by the AuNR (slope ~0), whereas the second peak position follows the PL (slope ~1.0). (B,D) Distribution of transition ratios, R, as a function of the PL position for a dye at Pos1 (B) and laterally displaced to its center, Pos2 (D). Blue shaded areas serve as a guide to the eye to compare the different cases. The gray dashed horizontal line is a reference marking the separation between enhancement (R < 1) or suppression (R > 1) of the secondary vibronic transitions with respect to the isolated molecule.*

Our findings reveal that the FL reshaping depends not only on the spectral overlap between plasmon resonances and molecular vibronic transitions, but also on the position of the dye, which sets the strength of their mutual coupling.[25] To further demonstrate this, we exploited the unique flexibility and nanometer positioning control of the DNA origami technique to place the dye laterally shifted ~25 nm from the AuNR tip towards the center (Pos2, see Figure S4). Exemplary spectra are shown in Figure S2. A reshaping is still observed, with the FL peak position also showing a linear dependence on the PL peak (Figure 3C, linear fit with slope ~1.01), although with higher dispersion. More importantly, the relative intensity of the peaks is now slightly different and the correlation with the PL peak position is a bit weaker than in the situation with the dye at the tip (Figure 3D). This is a sign of a weaker coupling between the AuNR and the dye molecule, as also evidenced from the lower total intensity collected compared to the reference (Figure S3).

To better understand the experimental results, we performed numerical simulations following the approach by Ringler *et al.*[22] In brief, this framework employs the Fermi Golden Rule to describe changes in the FL spectrum through the spectral variation of the radiative decay rate enhancement (radiative Purcell) $g_r(\lambda)$:

$$F(\lambda) \propto g_r(\lambda) F_0(\lambda) \tag{2}$$

where $F(\lambda)$ and $F_0(\lambda)$ represent the amplitude normalized FL spectrum of the dye in the presence and absence of the AuNR, respectively. Note that we have dropped spectrally-integrated quantum yield terms in Eq. 2,[22] as they are inherently wavelength-independent. Thus, to retrieve the FL spectrum, it is sufficient to simulate the radiative Purcell factor at the dye position (which gives the radiative decay rate enhancement that it experiences) and multiply it by the measured reference spectrum. The results for $g_r(\lambda)$ and $g_r(\lambda)F_0(\lambda)$, for a dye at Pos1 (inset of Figure 3A) and orientation along the longitudinal axis of the AuNR, are included in Figures 4A and 4B, respectively. In order to account for the size dispersion in our experiments, we employed for these simulations AuNRs with a diameter of 38 nm and different lengths. As expected, $g_r(\lambda)$ increases in intensity and redshifts as the length of the AuNRs is increased (Figure 4A). The product $g_r(\lambda)F_0(\lambda)$ (Figure 4B) also redshifts with the AuNR length, though less than $g_r(\lambda)$. Remarkably, the numerical predictions underestimate the FL reshaping induced by the AuNR, despite that the geometric and material idealizations inherent to local electrodynamics calculations overestimate the light-matter interaction strength in nanophotonic systems.[36,37] Note for instance that the fingerprint of the main peak of $F_0(\lambda)$ is apparent in all the spectra in Figure 4B. There is a number of factors that may be playing a significant role in the experiments, such as the dye-AuNR relative orientation and distance, or the size dispersion in the nanorod samples for a determined aspect ratio (see simulations in Figure S5). Our results indicate that none of them could induce an enhancement in $g_r(\lambda)$ large enough to reproduce the rich FL reshaping phenomenology in Figure 2. It is worth considering that Figure 4B corresponds to the dye-AuNR configuration that maximizes the Purcell effect in our calculations. In accordance with the experimental study (see Figure 3), we also analyze numerically the influence of the dye position on the FL spectrum next. The evaluation of Eq. 2 at Pos2 is included (dashed lines) in Figure 4C for three AuNR lengths. The comparison against the results for Pos1 (solid lines) reveals a similar plasmonic effect on the emission reshaping. This numerical result we can link to the spatially extended character of the bright, longitudinal mode behind the $g_r$ maxima in Figure 4A, which translates into a small variation of the radiative LDOS within the 25 nm distance between Pos1 and Pos2.

Overall, Figures 4B and 4C can rationalize the main features observed in the measurements included in Figures 2 and S2 in terms of the displacement of the FL peaks and their relative intensities, but do not reproduce the order of magnitude difference in the range of $R$ values between Figure 3B and 3D. The numerical calculations predict a weaker dependence of the FL reshaping on the dye position than observed in the experiments.

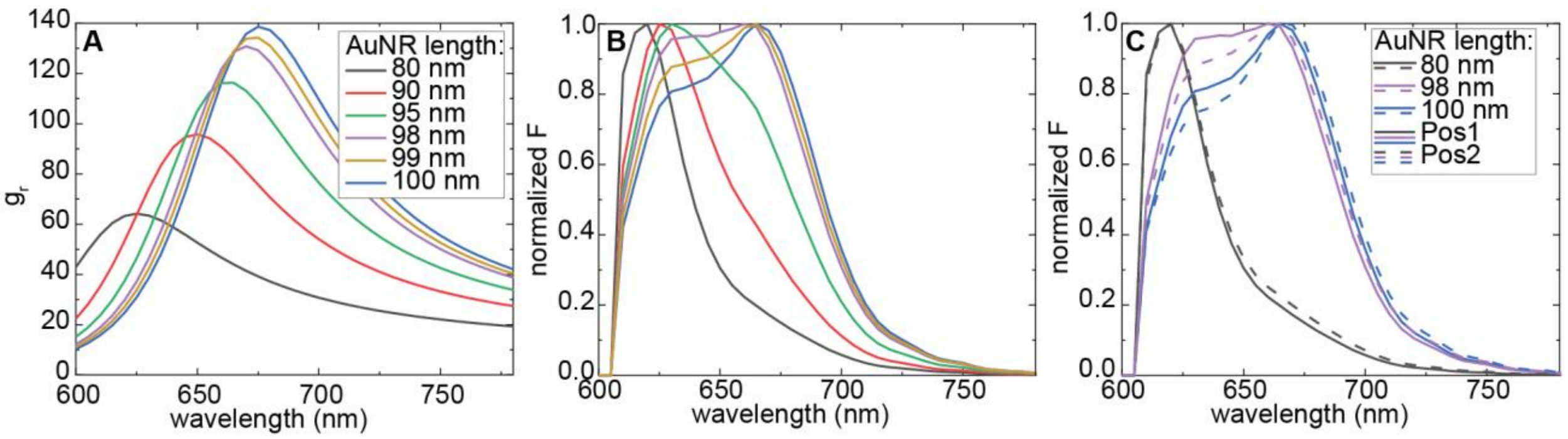

***Figure 4. Simulations of the spectral reshaping for dye-AuNRs of different lengths.*** *(A) Radiative Purcell $g_r(\lambda)$ experienced by a dye placed at Pos1 and oriented along the main axis of a AuNR of 38 nm in diameter and different lengths. (B) Corresponding spectral reshaping generated by the radiative Purcell factors simulated in (A). (C) Comparison between the spectral reshaping for dyes at Pos1 and Pos2.*

Using Eq. 2, we can experimentally estimate the spectral behavior of $g_r(\lambda)$ by dividing our FL data by the reference one. Examples corresponding to the spectra shown in Figure 2 are depicted in Figure 5. The comparison between the extracted $g_r(\lambda)$ (green) and the AuNR PL (orange) for Pos1 shows a close correspondence, although PL peaks are generally blueshifted from $g_r$ ones. This comparison resembles the observations recently reported for PL and scattering maxima,[34] although a closer match between PL and radiative Purcell spectra was expected, given that they share the same physical origin.[33]

A plausible explanation for the disagreement between the two sets of data in Figure 5 is the failure of Eq. 2 in our experimental samples, neglecting possible changes in dephasing due to the proximity of the nanorod. As discussed above, it assumes that the vibrational relaxation is much faster than the photon emission rate in the dyes. Note that this is implicit in the sketch of Figure 1A, where all the radiative transitions take place from the ground vibrational state of the excited electronic level, in accordance with Kasha's rule. This is a valid assumption in free space, where the timescale for the former (latter) is in the ps (ns) range,[21] three orders of magnitude apart. Our numerical simulations yield total Purcell factors (Figure S6) five times larger than $g_r(\lambda)$ in Figure 4A, thanks to the strong absorption by the AuNRs. This corresponds to a $\sim$700-fold reduction in the FL lifetime, making it comparable to vibrational decay. Furthermore, recent experiments using pump-probe techniques on single dyes at the hotspot of self-assembled optical antennas revealed fluorescence lifetimes smaller than 20 ps.[39] In this scenario, vibronic transitions from higher vibrational states in the electronic transitions, characterized by different Franck-Condon factors, take place. The theoretical description of these would require a theory describing photonic and vibrational degrees of freedom on the same footing,[20] but it is obvious that they will effectively alter $F_0(\lambda)$ in Eq. 2, which would no longer correspond to the free-space FL spectrum. Apart from rationalizing the discrepancies between PL and $g_r$ in Figure 5, this mechanism, strongly mediated by the total (instead of radiative) LDOS, would also explain why our numerical simulations underestimate the strength of the FL modification induced by the AuNR and its nanometric dependence on the emitter position. Figure S2 presents the comparison between experimentally extracted $g_r$ and PL for Pos2, showing even larger discrepancies (note however that we can expect plasmonic modes contributing to the former that are absent in the latter in this asymmetric configuration).

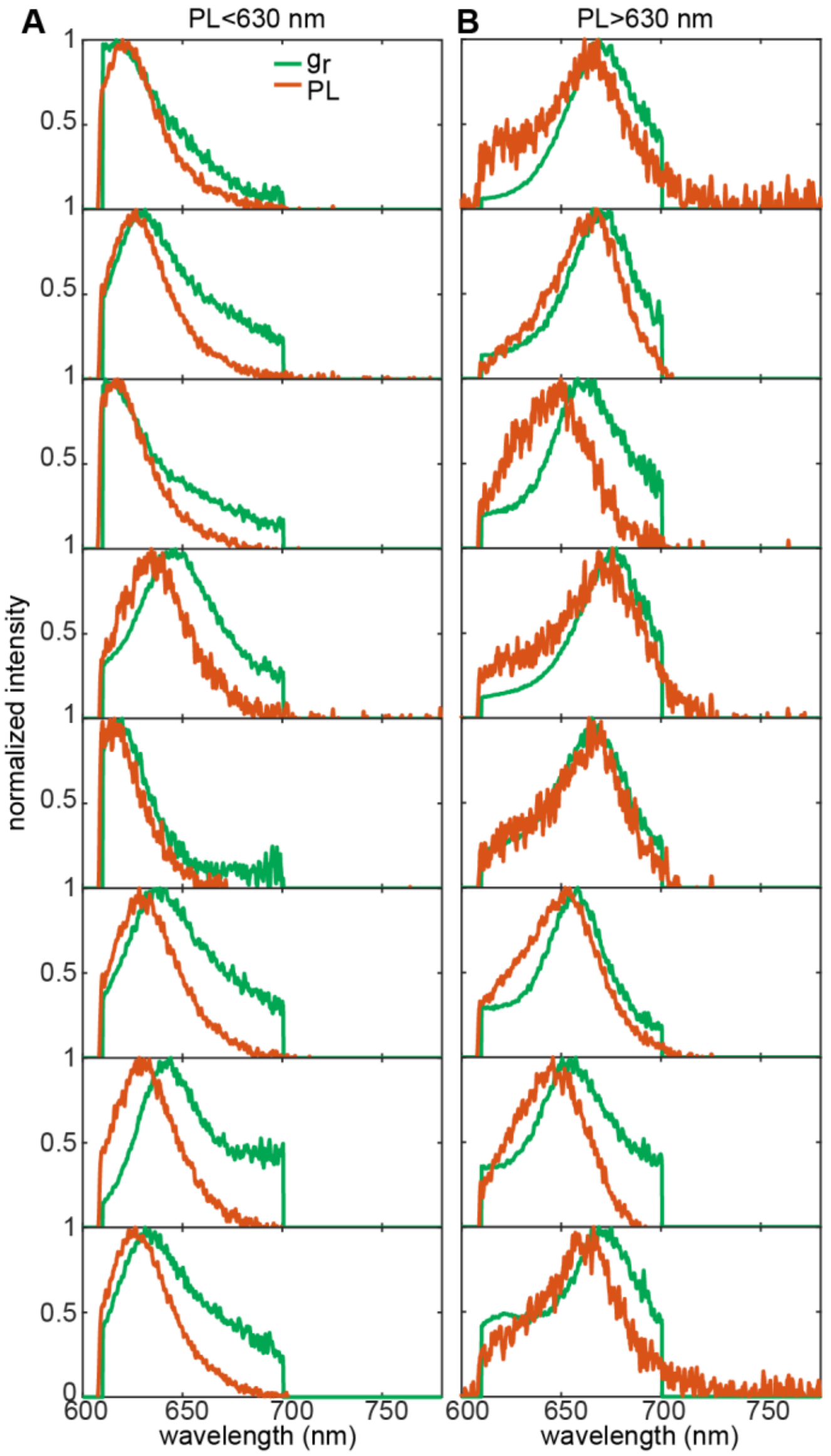

***Figure 5. Spectral mapping of radiative rate enhancement.*** *(A,B)* $g_r(\lambda)$ *retrieved from the spectra shown in Figures 2A and 2B, respectively, together with the corresponding AuNR PL spectra (both normalized to 1 for better comparison). Spectra are divided into cases with PL < 630 nm (A) and PL > 630 nm (B), i.e. in resonance with the main and secondary FL peaks of the isolated molecule.*

In summary, we have experimentally demonstrated reshaping of the emission spectrum of a single-molecule by means of coupling to a nanoantenna. We used bottom-up self-assembly of the dye-AuNR system by means of a T-shaped DNA origami structure, which allows us to precisely control their relative positions and thus their interaction. The main features of our results can be rationalized by numerical simulations and show that the decay rates to different vibrational sublevels of the electronic ground state can be controlled by matching them to the plasmon resonance of nearby optical antennas. Furthermore, laterally displacing the position of the dye with respect to the AuNR tip, we observe a lower degree of spectral reshaping, a sign of weaker coupling between both. Finally, we show that these changes in the fluorescence emission in presence of a nanoantenna can be used to measure the spectral behavior of the radiative decay rate enhancement produced by such nanostructure. We observe a discrepancy between the experimentally extracted radiative Purcell and the measured PL, pointing out to the existence of an additional mechanism for the decay. This can be explained by the high total Purcell factor of this configuration, yielding a FL lifetime that is comparable to vibrational decay and allowing vibronic transitions from higher vibrational states.

Overall, our observations indicate that, by tuning the spectral overlap between the intrinsic fluorescence of a dye and the resonance of a nanoparticle, one can basically design the emission spectrum of the system. In future works, we expect to obtain a higher degree of emission manipulation by controlling the dye's orientation[38] and by employing AuNRs with a reduced size dispersion.[42]

## Methods

### Numerical simulations

A frequency domain solver based on the finite element method in CST Studio Suite was used for the 3D full-wave simulations. The boundaries were set to “open added space” in the six faces of the simulation box. The single emitter was simulated by a discrete port with 5000 ohms combined with a dipole antenna of PEC material placed at 5 nm from the AuNR surface. AuNRs were modeled as cylinders (R = 19 nm, unless otherwise specified) with two semi ellipsoidal caps (a = b = R, c = 12 nm), and a length of 100nm (unless specified otherwise). For the dielectric function of gold, we used fitting data from the Lorentz–Drude model as in Elazar *et al.*[43]. To account for the DNA labelling of the AuNRs, a DNA coating layer of 2 nm was introduced with a refractive index of 1.7.[44] The radiative decay rate enhancement ($g_r = \gamma_r/\gamma_{r0}$) was calculated from the ratio of the radiated power of the dipole antenna in the presence and absence of the functionalized AuNR, $P_r$ and $P_{ro}$, respectively.[40,41]

### Experimental setup

Spectral measurements were performed at the exit port of a confocal microscopy body. A flip mirror was used to change between confocal imaging and spectral detection. Confocal imaging was first employed to identify the location of the structures. Then, the flip mirror was switched to direct the light from the single structures positioned at the center of the beam onto the spectrometer path. Emitted light was focused at the entrance slit of a spectrometer (Spectra Pro HRS 300, Princeton Instruments) coupled to a CCD (PIXIS:100B_eXcelon). For signal collection, a high NA objective was used (Olympus, 100× NA=1.4).

Excitation was performed with a supercontinuum white light laser (FYLA SCT1000) that was spectrally filtered to a wavelength of 594 nm (±5nm) to efficiently excite the dye (ATTO 594). Light was circularly polarized at the sample plane to achieve uniform excitation of all the structures. A long-pass filter at 610 nm was used on the collection to block laser excitation, causing the abrupt decay observed in some of the spectra (Figure 2 and S2).

**DNA origami design and folding**

The T-shape DNA origami was designed using CaDNAno[45] and visualized for twist correction using CanDo.[46] Figure S4 shows a cartoon with its dimensions and position of the modified staples. In short, the DNA template has Poly-A handles that serve to accommodate single AuNRs (Nanopartz INC, average size 40x80 nm), with 18x Poly-A8 (blue) and 2xPoly-A12 (green) ssDNA strands (Biomers.net GmbH) on the right side. Single ATTO 594 dye (red, Biomers.net GmbH) were positioned at the top of the T-design (at Pos1 or Pos2). The design files were uploaded at https://nanobase.org/.[47] Unmodified DNA sequences were purchased from Integrated DNA Technologies, INC. Folding was performed based on the following protocol: a scaffold consisting of a vector derived from the single-stranded M13-bacteriophage genome (M13mp18, 7249 bases, Bayou Biolabs) and staples (100 nM, ca. 32 nts) were mixed in a 1x TAE buffer (40 mM Tris, 10 mM Acetate, 1 mM EDTA, pH 8, stock purchased from Alfa, CAS#77-86-1, J63931.k3) containing 12 mM $MgCl_2$ (stock purchased from Alfa, CAS#7786-30-3, J61014.AK). The solution was heated to 75°C and ramped down to 25°C at a rate of 1 degree every 20 mins. The folded DNA origami structures were purified from excess staples strands by gel electrophoresis using a 0.8 % agarose gel (LE Agarose, Biozym Scientific GmbH) in a 1x TAE buffer/12 mM $MgCl_2$ buffer for 2.5 hours at 4 V/cm. The appropriate band containing the targeted DNA template was cut out and squeezed using coverslips wrapped in parafilm.

**Nanorods functionalization and attachment to the DNA template**

Thiolated DNA (Thiol-C6-T18, ELLA Biotech GmbH) was mixed with AuNRs at a 100 nmol:100 uL@100OD ratio and frozen overnight.[48] Excess DNA was removed using gel electrophoresis. This step also ensures the removal of any self-aggregated dimer formed during the AuNR functionalization. The concentration was determined via UV-Vis absorption spectroscopy (Nanodrop).

The purified DNA template was mixed with the purified AuNRs using an excess of five NPs per binding site. After overnight incubation at room temperature, the excess of AuNRs was removed by gel electrophoresis (running for 4.5 hours) and the band containing correctly formed structures (single AuNRs) was extracted as described before.

**Surface Preparation and Sample Immobilization**

For immobilization of the structures, glass cover slips were first sonicated in consecutive baths of acetone, propan-1-ol and water for 5 minutes each, and then cleaned in a UV cleaning system (PSD Pro System, Novascan Technologies, USA). The surfaces were then passivated with a BSA-biotin/PBS solution (0.5 mg/mL, Sigma Aldrich, CAS#9048-46-8, A8549-10MG) for 20 mins, neutrAvidin/PBS solution (0.5 mg/mL, Thermofischer, 10443985) for 20 min, and 5’ biotin-Poly-A15-3’ ssDNA (Biomers.net GmbH). In between all the steps, the surfaces were washed with 1x PBS buffer (Alfa, J75889.K2). The samples were incubated on the surfaces for 2h.

**Supplementary figures**

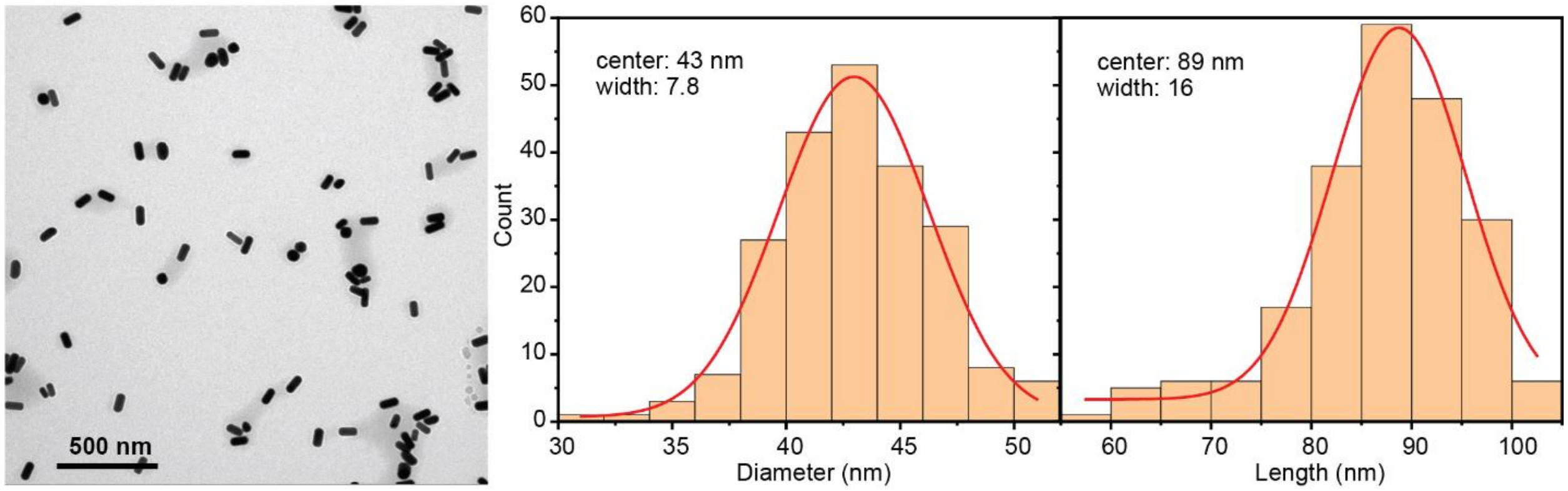

**Figure S1. Size dispersion of AuNRs.** TEM image of the AuNRs used in this work, showing the size variability, together with histograms of the diameter and length distribution of the AuNRs. Insets display center and width of the Gaussian fits.

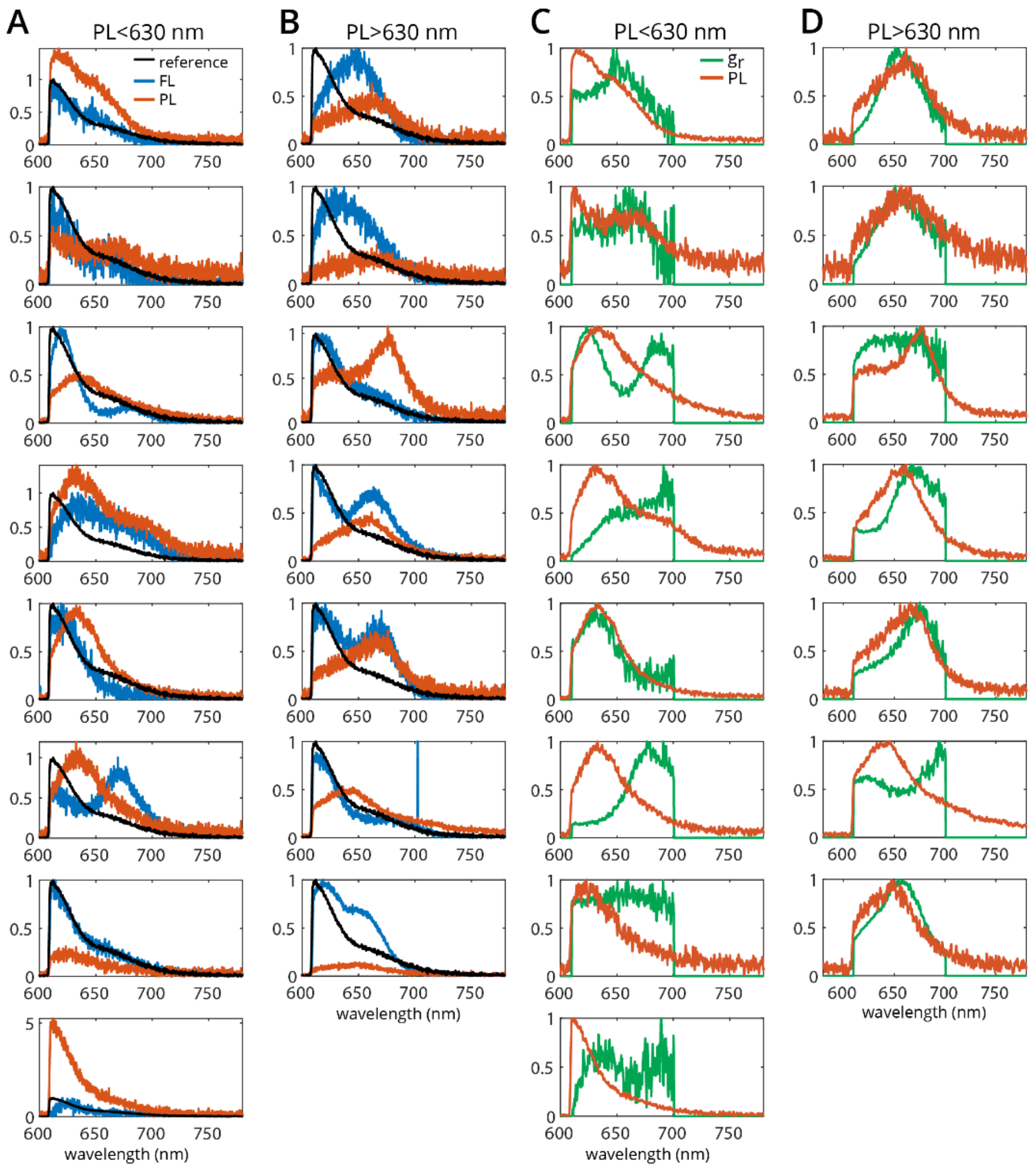

**Figure S2. Reshaped fluorescence spectra from structures with different PL for a dye at Pos2.** (A,B) Fluorescence emission spectra for the dye without AuNR (reference, black line) and for the hybrid dye-AuNR structures (dye at Pos2). The AuNR PL is also included for each case (orange line). Spectra are divided into cases, with PL < 630 nm (A) and PL > 630 nm (B), *i.e.* in resonance with main and secondary (shoulder) FL peaks. PL spectra are normalized to the corresponding FL maxima for visualization of their intensity difference. (C,D) $g_r(\lambda)$ retrieved from the spectra in (A,B), together with the corresponding AuNR PL spectra (both normalized to 1 for better comparison).

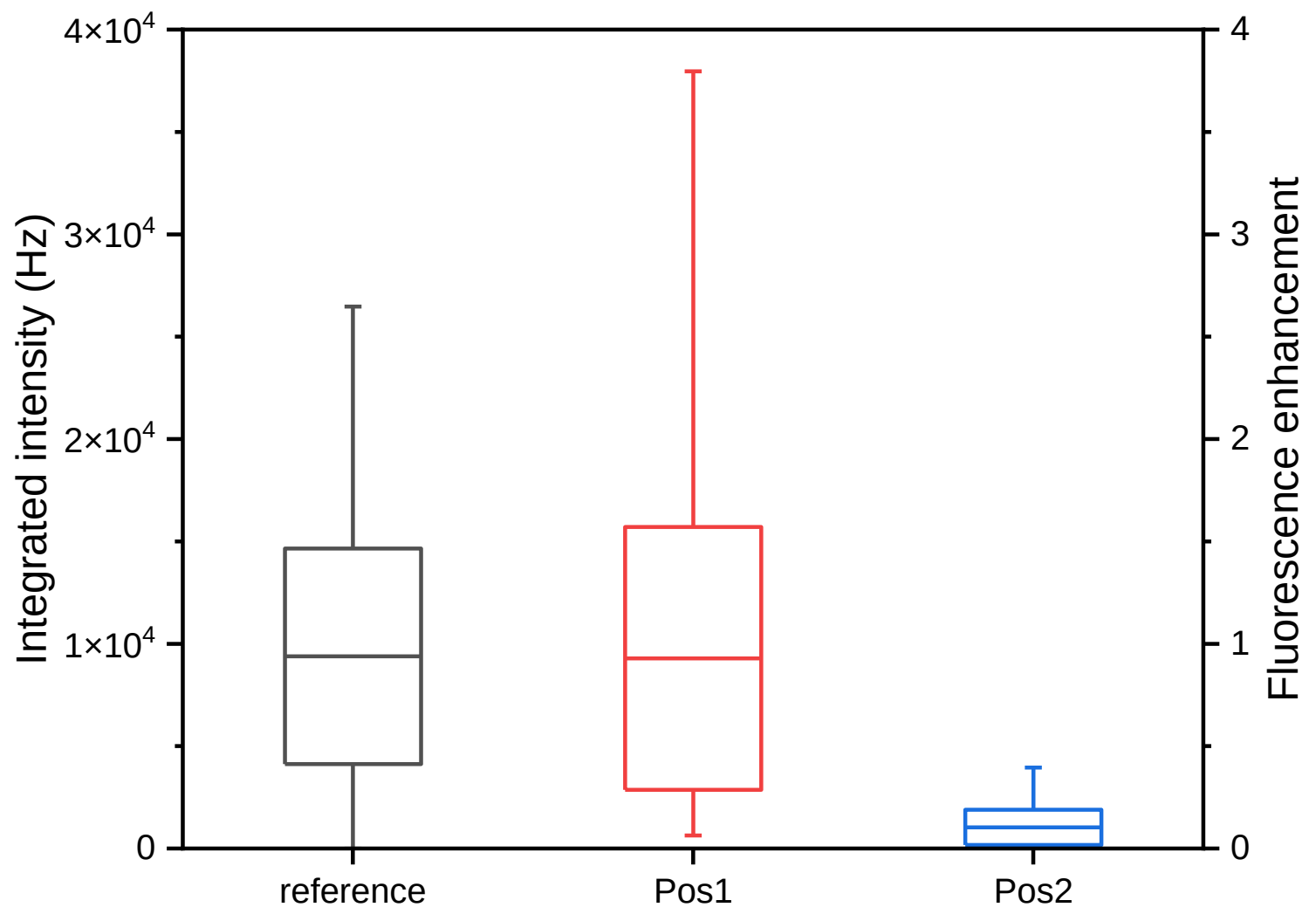

**Figure S3. Total FL intensity for the different positions of the dye around the AuNR.** Distribution of the integrated FL intensity in the whole spectral range for all the measured spots for the origami in absence (reference) and presence (Pos1 and Pos2) of the AuNR. Lines represent minimum and maximum values and the box shows the mean with the corresponding standard deviation. As expected, differential coupling to the AuNR is observed for the various potions of the dye around it.

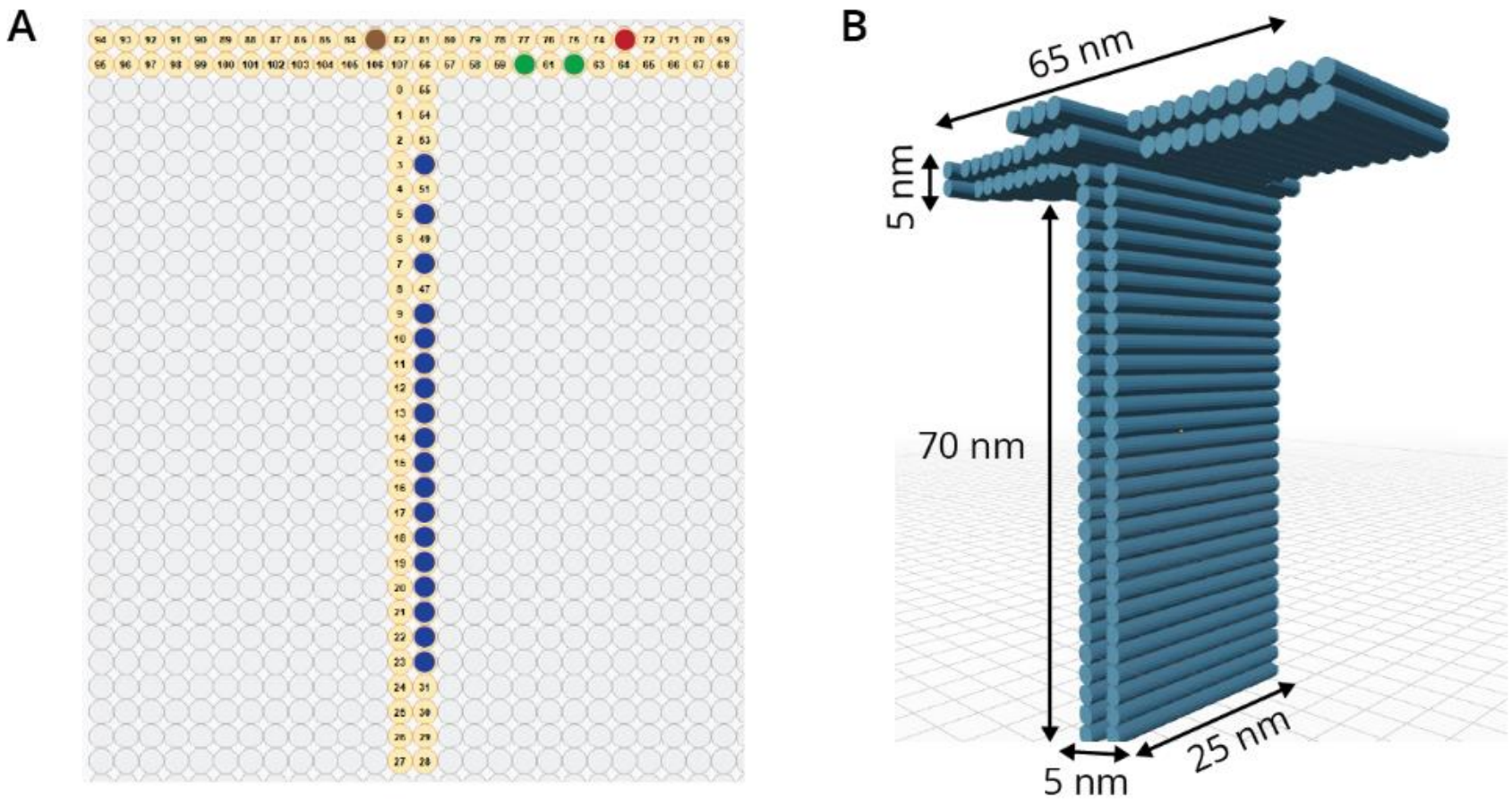

**Figure S4. T-shape DNA origami design.** (A) caDNAno Bundle view of the T-shape DNA template. Poly-A handles (A8 in blue and A12 in green) serve to assemble a single 40×80 nm AuNR. The DNA-template can accommodate a single ATTO 594 fluorophore either at Pos1 (red) or at Pos2 (brown) to form the different configurations as explained in the main text. (B) 3D cartoon of the DNA origami indicating dimensions.

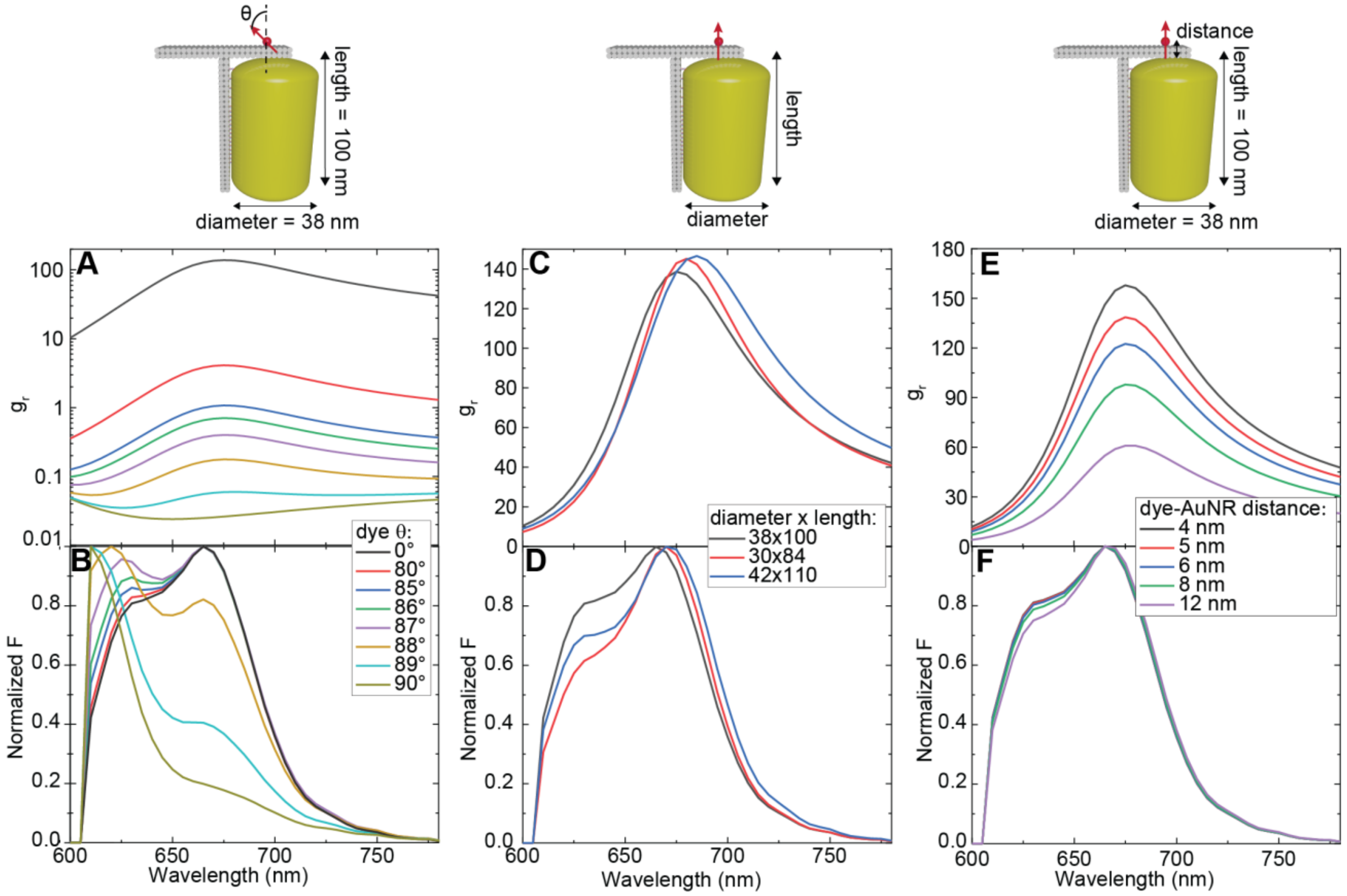

**Figure S5. Effect of dipole orientation, distance and AuNR dimensions.** (A,C,E) Simulated $g_r(\lambda)$ and (B,D,F) corresponding spectral reshaping for a dye at Pos1 (A,B) at different orientations, (C,D) coupled to AuNR of different dimensions having the same resonance wavelength, and (E,F) placed at various axial distances from the AuNR tip. At the top of each row, a sketch is shown depicting the parameters utilized for each situation.

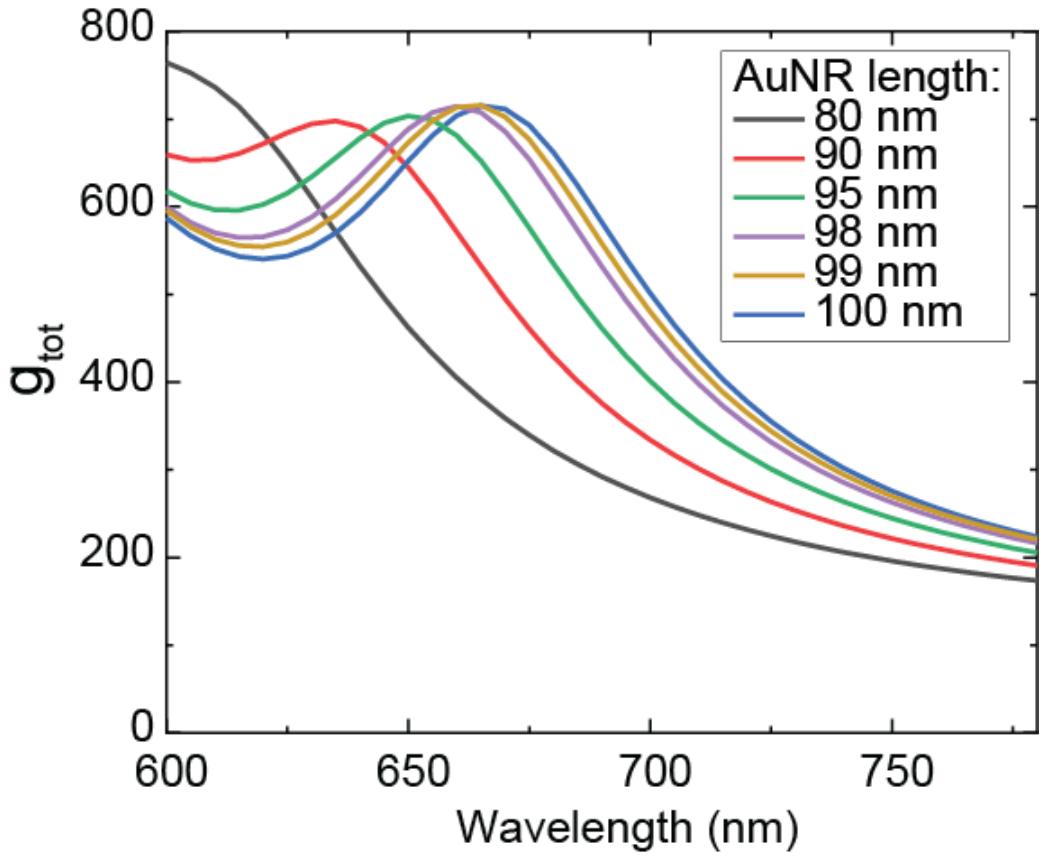

**Figure S6. Simulations of total Purcell factor.** Total Purcell factor $g_{tot}(\lambda)$ experienced by a dye positioned at Pos1 and oriented along the main axis of a AuNR of 38 nm in diameter and different lengths.